**Light-driven lattice metastability for enhanced superconductivity in FeSe/SrTiO$_3$**

Qiang Zou[1], Zhan Su[2], Andres Tellez Mora[1], Na Wu[2], Joseph Benigno[1], Christopher L. Jacobs[1], Aldo H. Romero[1], Subhasish Mandal[1], Yaxian Wang[2], Sheng Meng[2], Michael Weinert[3], Hua Zhou[4,*], Lian Li[1,*], Cheng Cen[1,2*]

[1]*Department of Physics and Astronomy, West Virginia University, Morgantown, WV, 26506, USA.*
[2]*Beijing National Laboratory for Condensed Matter Physics, Institute of Physics, Chinese Academy of Sciences, Beijing 100190, People's Republic of China.*
[3]*Department of Physics, University of Wisconsin, Milwaukee, Wisconsin 53211, United States.*
[4]*X-ray Science Division, Advanced Photon Source, Argonne National Laboratory, Lemont, IL, 60439, USA.*

(Dated: April 24, 2025)

Driven quantum materials with on demand properties controlled by external stimuli are critical for emergent quantum technology. In optically tunable superconducting heterostructures, the lattice responses at the buried interface may hold the key to the light susceptibility but is very challenging to detect. In this work, a nondestructive synchrotron-based X-ray scattering phase-retrieval technique is implemented in monolayer-FeSe/SrTiO$_3$ heterostructures to capture the three-dimensional interfacial atomic displacements in-situ as the interface superconductivity is actively manipulated by light. It is found that the interlayer sliding between FeSe and SrTiO$_3$ can drastically alter how the lattice responds to the light. In domains with selected stacking configurations, the interface transforms the very weak photoexcitation in SrTiO$_3$ into significant Fe-atom displacements in FeSe and generate metastable interfacial structures that can lead to a persistent superconductivity enhancement. These findings demonstrate an effective strategy for achieving greatly amplified light-lattice coupling for efficient quantum phase manipulations at designed interfaces.

# 1. Introduction

Light has long been an effective tool for controlling the electronic and spintronic properties in solid-state systems for emergent functionalities.[1-5] More recent discoveries showed that correlated quantum phases can also be tailored by directly manipulating the lattice structure using light.[6] For example, ultrafast photoexcitation is capable of triggering transient interlayer sliding in WTe2 and producing a transition from a centrosymmetric quantum spin hall insulator to a type-II Weyl semimetal.[7] Another example is in MoTe2, where strong laser irradiations can generate defects and induce a transition from a hexagonal semiconductor to a monoclinic metal with greatly improved carrier mobility through electrical contacts.[8]

Nevertheless, these works often need to counteract the typically very weak light-lattice interaction by either generating an enormous transient field strength impulsively [7, 9-11] or relying on large power intensities to modify the lattice through time-accumulated effects such as heating or chemical reactions [8, 12-13].

Far more efficient optical control over the lattice is possible at designed heterostructures, where the coupling between two distinct materials at the interface can greatly amplify the effects of light. In the epitaxial heterostructure between FeSe monolayer and bulk $SrTiO_3$(001) substrate, the interface-enhanced superconductivity has exhibited a superconducting transition temperature ($T_C$) almost ten times higher than the bulk.[14-15] Earlier work [16] also found that the $T_C$ can be swiftly and persistently raised by very weak continuous wave (CW) ultraviolet (UV) photoexcitations with power intensities well below 10 µW/cm$^2$. It was suggested that such an effect may result from a light-induced structural transition that is highly localized at the interface. This mechanism, however, has been difficult to verify due to the lack of tools that can effectively detect atomic displacements at buried interfaces without introducing extrinsic modifications.

To address this challenge, we implemented a unique synchrotron surface X-ray scattering technique to capture the three-dimensional (3D) atomic displacements at the interface with a sub-Å resolution. This technique, crystal truncation rod (CTR), takes advantage of the superior brightness and coherence of synchrotron sources to allow the detection and analysis of the weak X-ray diffraction variations caused by the broken lattice periodicity at interfaces and surfaces.[17-21] Employing coherent Bragg rod analysis (COBRA), an iterative phase-retrieval technique, a complete 3D electron density map of the interface can be reconstructed from a group of experimentally measured CTRs positioned differently in the reciprocal space,[22-24] uncovering the changes in the chemical distribution, lattice structure, and charge ordering going from the bulk to the interfacial layers.[21]

Using integrated UV light sources inside the CTR sample chamber, we were able to directly compare the 3D electron densities near the FeSe/SrTiO$_3$ interface before and after weak CW photoexcitation. Profound light-induced metastable lattice distortions in the FeSe monolayer at low temperatures were detected. The nature of these structural changes highly depends on how FeSe is stacked on top of the SrTiO$_3$ substrate. The most significant out-of-plane atomic displacement, as large as 1/3 of the monolayer thickness, occurs in the regions where Fe atoms are aligned directly on top of the Ti atoms at the SrTiO$_3$ surface. Sliding the FeSe layer by half a unit cell, the light-induced displacements become dominantly in-plane. Altering the spectral weights in Fe 3$d$ orbitals and modifying the electron-phonon coupling strength, these actively controllable lattice distortions can strongly contribute to the $T_C$ tuning.[25] The overall results not only demonstrate a promising venue for actively manipulating the lattice of 2D quantum materials using weak lights but also experimentally identify structural characteristics that the interface superconductivity is particularly susceptible to, suggesting possible directions for future studies to better understand the interface-assisted pairing mechanism for high-temperature superconductivity in single-layer FeSe/SrTiO$_3$.

## 2. Results

### 2.1. CTR measurements with in-situ photoexcitation

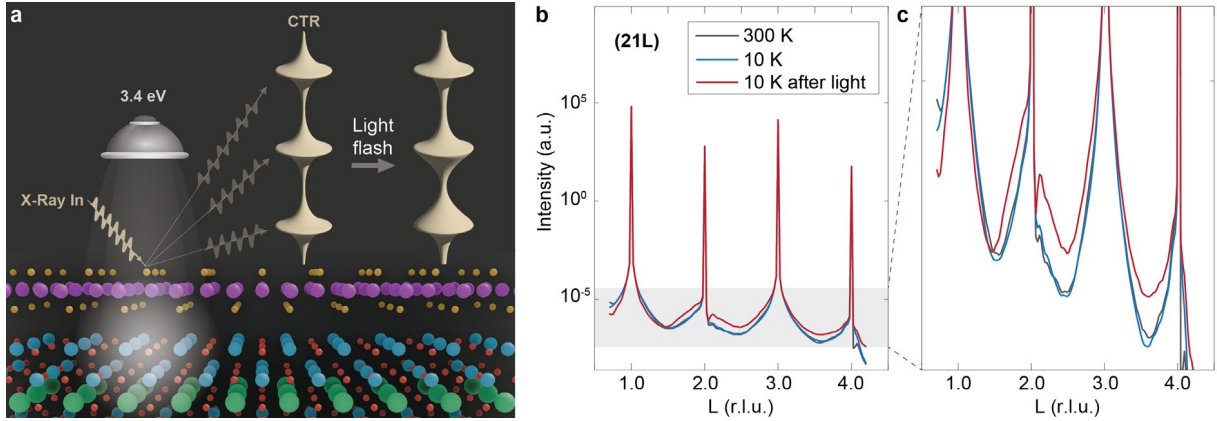

**Figure 1 Light-induced interfacial structural transition detected by CTR-COBRA technique (a)** Examining the X-ray diffraction signals away from the reciprocal lattice Bragg points as well as their changes in response to the 3.4 eV light excitation, the CTR measurements characterize the light-induced atomic displacements at the interface where the bulk lattice periodicity is broken. **(b)** Typical sets of CTR data measured at in-plane reciprocal coordinates (2,1). The index L represents the position change along the out-of-plane axis in the reciprocal space. **(c)** Magnified view of the middle-zone X-ray diffraction intensity between the bulk peaks (shaded by grey in (b)), which changes significantly before and after the light excitation.

Epitaxially synthesized FeSe monolayer on insulating $SrTiO_3$ substrate was capped by a protective layer before transferring to the synchrotron facility for CTR measurements. To avoid the X-ray signal interference caused by the capping layer, the capping materials used in this study are either amorphous or crystalline but have a lattice symmetry distinct from FeSe. We focus on the sample capped by crystalline $Sb_2Te_3$ in the main text, and the data from the amorphous Se-capped sample showing consistent results is shown in the supplementary information (Figure S5, S6 and S7, Supporting Information).

Figure 1b shows a typical CTR data set along the (21L) rod. Here, the indices inside brackets represent the in-plane reciprocal coordinates of the CTR measurements. Besides the bulk X-ray diffraction peaks from the $SrTiO_3$ substrate, the orders-of-magnitude weaker off-peak diffraction signals generated by locally broken lattice periodicity are also clearly resolved (Figure 1c). After the sample is illuminated by a 3.4 eV light emitting diode (LED) for a couple of seconds, the bulk peaks remain the same, but the signals along the CTR between the bulk peaks vary significantly. These features found in a group of CTR rods are later processed by COBRA analysis and found to be produced by the lattice distortions in the FeSe monolayer. Before further elaborating on these phenomena in the following texts, we note that no change in the CTR data was detected when a white light LED covering the 4000 K blackbody radiation spectrum (photon energy range of 1.6-3.2 eV) was used instead. The presence of a photon energy threshold near 3.2 eV (*i.e.*, the bandgap of $SrTiO_3$), consistent with what was found in the previous transport measurements,[16] indicates that the photoexcitation responsible for the structure distortions in FeSe primarily takes place in the $SrTiO_3$ substrate.

## 2.2. Domains formed from the different epitaxial relations at the interface

The reconstructed 3D electron density (Figure 2a) shows a modified lattice structure near the interface compared to the bulk. Along the out-of-plane [001] direction, the $SrTiO_3$ bulk is composed of alternating SrO and $TiO_2$ layers and terminated by a $TiO_x$-$TiO_2$ double-layer.[19, 26-27] The FeSe monolayer is built from a Fe layer sandwiched by two Se layers. Unlike the bulk structure, where the three layers are equally spaced, the Fe layer in our samples is always slightly closer to the bottom Se layer (Figure S1, Supporting Information), possibly due to the asymmetric interactions with the substrate (Figure S2, Supporting Information). An additional layer of Se with a much lower density is found between FeSe and $SrTiO_3$, which is also seen in TEM results obtained on our FeSe films grown on conducting $SrTiO_3$ substrates [28] but is, in this case, placed closer to the FeSe monolayer. This overall layer-by-layer configuration near the interface is mostly the same in all samples tested, independent of the different capping materials used.

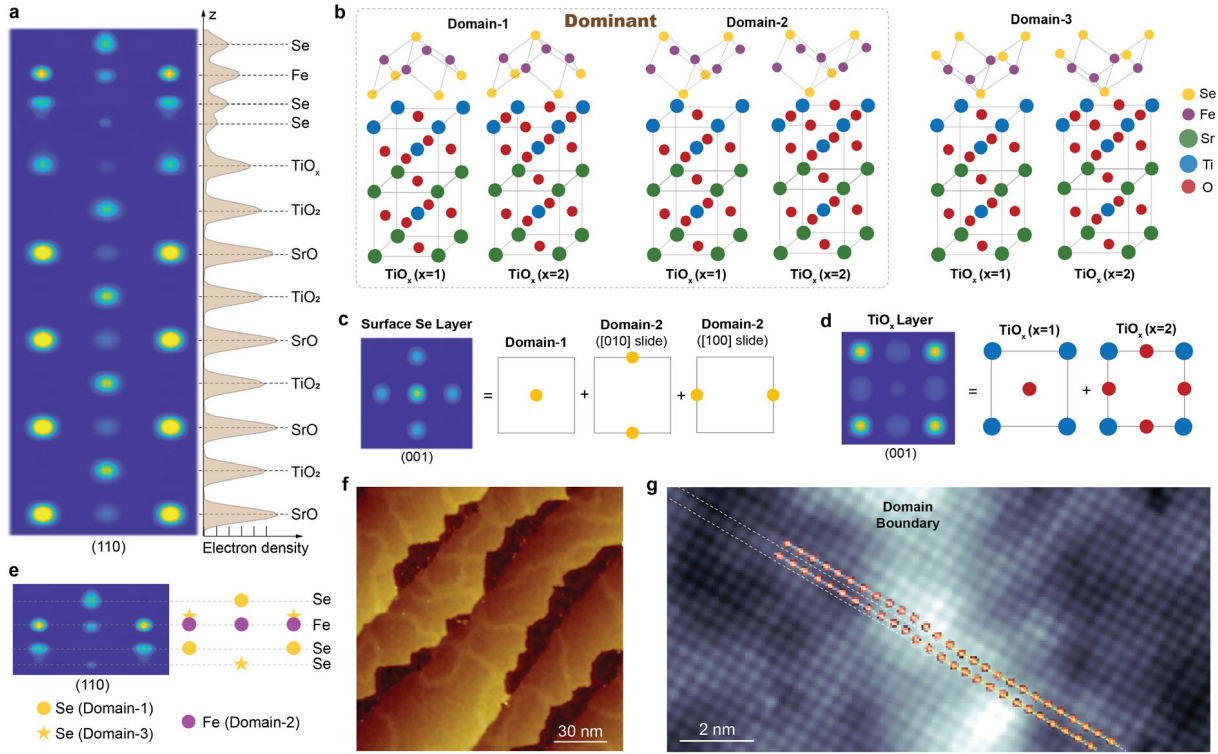

**Figure 2 3D atomic structure of the FeSe/SrTiO₃ interface (a)** The layer-by-layer structure of the interface represented by the (110) atomic plane cut of the 3D electron density retrieved by COBRA analysis and the depth-dependent plot calculated by integrating the signal in each (001) atomic plane. Different color scales are used for the FeSe and SrTiO₃ sections to simultaneously visualize atoms with very different atomic weights. **(b)** Illustration of the six possible FeSe/SrTiO₃ interface structures formed from the three FeSe domains (Domain-1, 2, 3) and the two types of TiO$_x$ surface reconstructions (TiO$_x$ ($x$ =1, 2)). Among them, interface configurations formed from Domain-1 and 2 are found to be dominant in our samples. **(c)** The electron density in the top Se layer is plotted as an example to show that the CTR-COBRA results point to a superposition of two domains that are related by a half-unit cell in-plane sliding of the FeSe layer. **(d)** Electron density in the TiO$_x$ substrate termination layer represents a superposition of two structures with different oxygen contents ($x$ = 1 or 2) and positions. **(e)** Identifications of the atoms in the FeSe layer that are visible in the (110) atomic plane electron density image displayed in (a), showing the possible presence of Domain-3 structure, with a low occupation density though, that is placed closer to the SrTiO₃ substrate. **(f)** STM image of a FeSe monolayer sample synthesized on a conducting SrTiO₃ substrate, showing the formation of the domain structures after post-annealing. **(g)** Magnified STM image near the domain boundary, where the ½ uc lattice offset between two adjacent domains can be clearly seen.

CTR measurements reveal the presence of multiple FeSe stacking domains (Figure 2b). Domain-1 corresponds to the stacking configuration where the atoms in the bottom Se layer is aligned with the Ti atoms in the topmost TiO$_x$ layer. Domain-2 differs from Domain-1 by a 1/2-unit cell slide of FeSe along either [100] or [010] axis. Different from local measurement tools such as STM, CTR characterizes the interface structures across the whole sample, which thus yields 3D electron density maps superposing the contributions from different domains (*i.e.* folded electron density maps from different structural domains). As an example, Figure 2c shows the electron density in the topmost Se layer. The bright spot at the square center is associated with Se atoms in Domain-1, and the four less bright spots at the edge centers are from Se atoms in the two degenerate Domain-2 structures. Besides these two dominant domains, a 1/2-unit cell translation of Domain-1 FeSe unit cell along the $[110]$ or $[\bar{1}10]$ axis can also form a third possible domain (Domain-3). This stacking configuration, however, is found to be very rare in our samples, as consistently indicated by both CTR and ARPES measurements (Figure S3, Supporting Information). If present, the FeSe monolayer in Domain-3 might be located much closer to the $SrTiO_3$ substrate, and contributes to the additional Se layer detected between FeSe and $SrTiO_3$ (Figure 2e). The co-existence of different domains is also seen in our samples grown on conducting $SrTiO_3$ substrates (Figure 2f), where the half-unit cell in-plane shift of the FeSe lattice can be directly visualized by scanning tunneling microscope (STM) at the domain boundary (Figure 2g).

A variation in the $SrTiO_3$ surface reconstruction is also observed. As shown in Figure 2d, the electron density of the $TiO_x$ surface layer (*i.e.* the outer layer of $TiO_2$ double-layer) represents the superposition of two different oxygen arrangements, one corresponding to $x = 1$ and the other to $x = 2$. These two $TiO_x$ structures plus the two dominant FeSe stacking domains form four most likely interface configurations (Figure 2b, inside dashed box). The electronic band structures of the four configurations, as computed by density functional theory (DFT), are somewhat similar, and all consistent with the angle-resolved photoemission spectroscopy (ARPES) results obtained before the deposition of the capping layer (Figure S3, Supporting Information).

## 2.3. Light-induced metastable structures in different FeSe domains

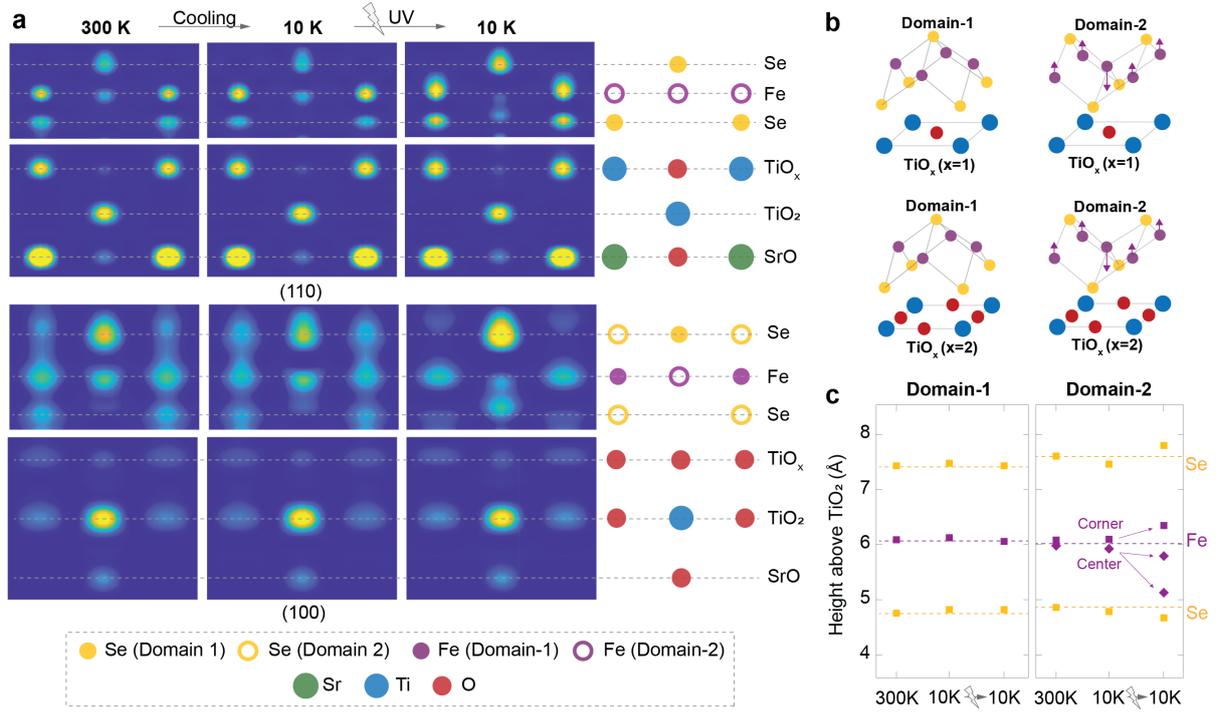

**Figure 3 Light-induced out-of-plane atomic displacements (a)** (110) and (100) atomic plane 2D view of the 3D electron density in the FeSe monolayer and the top three atomic layers in SrTiO$_3$, measured at 300 K, 10 K before the light excitation and at 10 K after the light excitation. Different color scales are used for the FeSe and SrTiO$_3$ sections to simultaneously visualize atoms with very different atomic weights. **(b)** Illustration of the domain-dependent out-of-plane displacements of Fe induced by light. In Domain-2, the Fe layer becomes highly corrugated after the light excitations. In Domain-1, the $z$ positions of the Fe atoms remain essentially unchanged. **(c)** The heights of the atoms in the FeSe monolayer above the topmost bulk TiO$_2$ layer in SrTiO$_3$. After the light excitation, the center-site Fe in Domain-2 occupies two positions separated by a height difference larger than 0.7 Å.

The domain-dependent lattice structure at the interface is significantly modified by photoexcitation at low temperatures. We first discuss the out-of-plane atomic displacements induced by light (Figure 3). As the sample is cooled from 300 K to 10 K, only a subtle variation of the interface structure appears. Nonetheless, an abrupt change occurs in the FeSe monolayer after the photoexcitation at 10 K (Fig. 3a). A closer examination of the data shows that such change mainly takes place in Domain-2. After the photoexcitation, Fe atoms in Domain-2 that are aligned on top of Ti (corner-Fe) move upward away from the interface. At the same time, the other Fe atoms (center-Fe) move downward, leading to a highly corrugated Fe-plane in Domain-2 (Fig.3b). The displacement of the center-Fe toward SrTiO$_3$ is particularly substantial, which also exhibits a spatial variation in their light-modified *z*-positions. The center-Fe atoms in some regions move 0.1 Å away from their original position. In the other regions, they move more than 0.8 Å closer to the interface (Figure 3c). Such twofold spatial variation may be correlated with the two types of TiO$_x$ surface reconstructions observed, which, however, cannot be conclusively verified by the CTR data alone. On the other hand, the out-of-plane displacements of the atoms in Domain-1 are very minimal, as are the atoms in the SrTiO$_3$ interfacial layers.

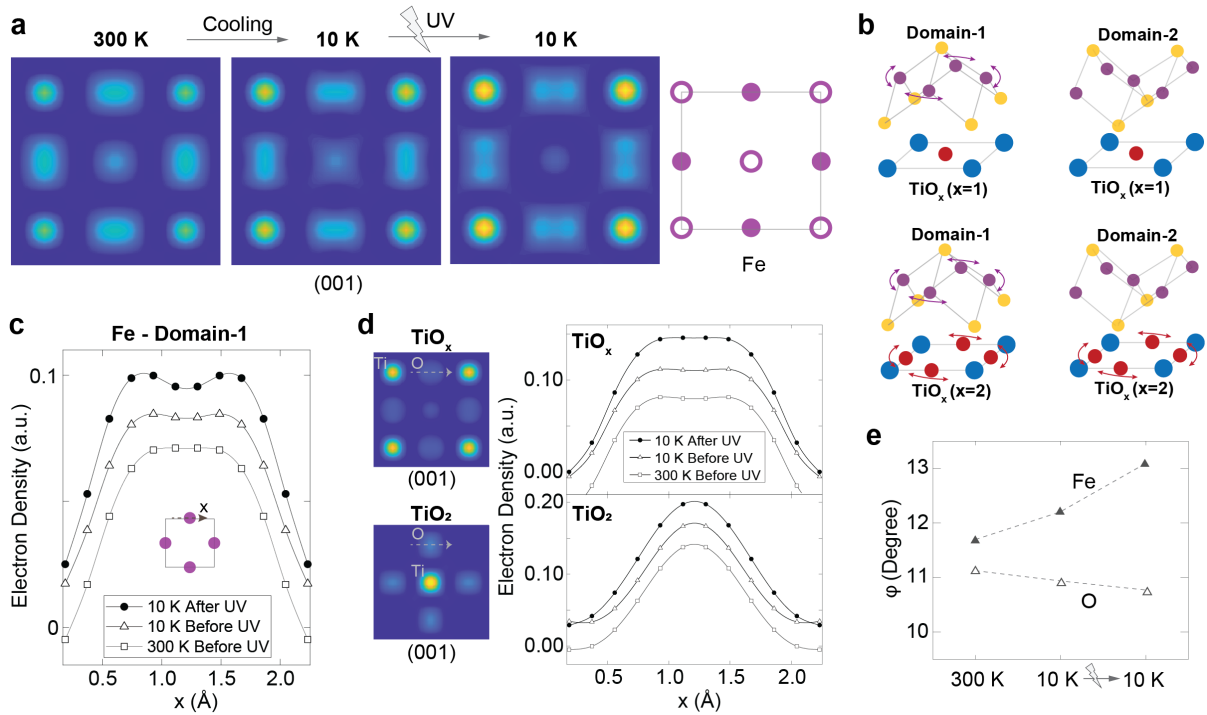

**Figure 4 Light-induced in-plane atomic displacements (a)** (001) atomic plane view of the Fe layer electron density measured at 300 K, 10 K before the light excitation and at 10 K after the light excitation. **(b)** Illustration of the in-plane rotations of the Fe- or O- trapezoids induced by light (rotation angle φ), which are only observed in FeSe Domain-1 and the TiO$_x$ ($x = 2$) region. **(c)** [100] direction line cut of the electron density near the Fe atom in Domain-1, indicating an in-plane shift of the Fe atoms in Domain-1 that becomes more significant after the light excitation. **(d)** Image and linecut plots of the electron density in the interfacial TiO$_x$ layer and the neighboring TiO$_2$ layer, showing the displacements of the oxygen atoms in the TiO$_x$ ($x = 2$) structure. **(e)** In-plane displacements of the Fe atom in Domain-1 and the O atom in TiO$_x$ ($x = 2$) region quantified by the equivalent rotation angles (φ) of the Fe- and O- trapezoids.

We then evaluate the in-plane atomic displacements by examining the electron density changes in the (001) atomic planes (Fig.4). In the Fe layer, the atoms in Domain-1 are offset from their high-symmetry position (Fig.4a), which likely comes from the antiferrodistortive rotations of the Fe-trapezoid (Fig.4b). From 300 K to 10 K, this rotation angle increases by $\Delta\varphi = 0.5°$ (Fig.4e). After the photoexcitation, such rotation gets abruptly enhanced by 1°. Similar in-plane displacement is also seen in the $TiO_x$ ($x = 2$) layer, though with an opposite trend. The O atoms in $TiO_x$ ($x = 2$) move closer to their high-symmetry position as temperature reduces, and photoexcitation moves them further in the same direction (Figure 4e). In comparison, no in-plane displacement is observed for atoms in Domain-2 FeSe or in $TiO_x$ ($x = 1$) structure.

**2.4. Links between the photoexcitation in SrTiO3 and the structural changes in FeSe**

As already discussed, $SrTiO_3$ is responsible for the absorption of the 3.4 eV photons. However, as shown in Figure 3 and 4, the most profound light-induced structural distortions take place in the FeSe monolayer. It is natural to ask how these two effects are linked microscopically. We now look at the photoresponses in the $SrTiO_3$ surface layers. As already shown in Figure 4d, O atoms at the $SrTiO_3$ surface inside the $TiO_x$ ($x = 2$) region undergo antiferrodistortive in-plane displacements that are reduced by light. Fig.5a plots the layer-specific out-of-plane polarizations in $SrTiO_3$, going from the bulk (left) to the interface (right). This value is calculated by first extracting the relative displacement between the anion and cation in each layer from the 3D electron density data and then multiplying it with the nominal ion charge. The polarization in the bulk layers is very small but becomes significant near the interface, especially in the $TiO_x$-$TiO_2$ double layer. The Ti atoms in the $TiO_x$ layer protrude closer to the interface than the O atoms in the same layer, thus producing a positive polarization along the [001] axis. The Ti atoms in the adjacent $TiO_2$ layer fall closer to the bulk, generating a negative polarization. Despite small variations, this overall polarization distribution is the same before and after the light excitation. The presence of structural polarizations generates a non-constant electrostatic potential profile inside SrTiO$_3$ (Figure 5b). This potential first turns more negative going from the bulk toward the interface, then, it is sharply neutralized by the large positive polarization in the TiO$_x$ layer, forming an electron potential barrier separating the bulk SrTiO$_3$ and the interface.

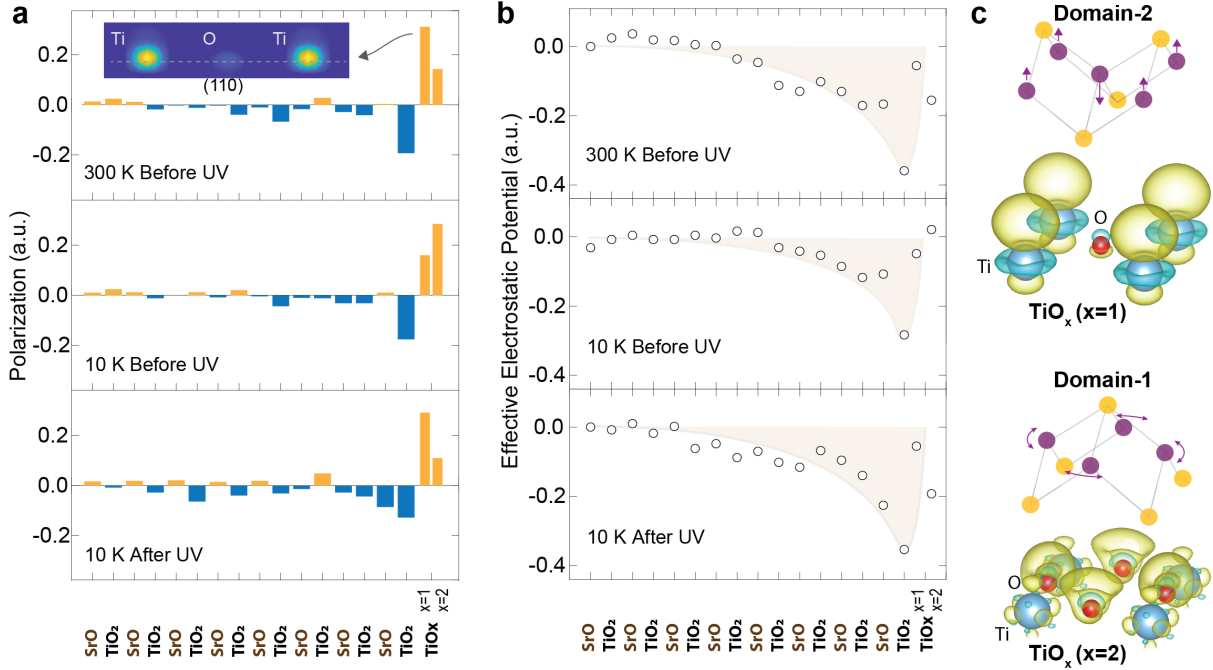

**Figure 5 Photoresponses of the SrTiO$_3$ surface layers (a)** Polarizations of each SrTiO$_3$ atomic layer calculated by multiplying the relative displacement between the metal and oxygen atoms and the normal charge of the ions. A sizable upward polarization is present in the top TiO$_x$ layer, screened by the downward polarizations in the neighboring TiO$_2$ and SrO layers that decay toward the bulk. The inset in (a) shows the (110) atomic plane electron density in the TiO$_x$ layer where the relative offset between Ti and O in the *z* direction is visually apparent. **(b)** An effective electrostatic potential is calculated by integrating the polarization of each layer with respect to their *z* positions, yielding an electron potential barrier near the interface. **(c)** Isosurface plot of the computed charge distribution change induced by photodoping in SrTiO$_3$ with different surface terminations. Also illustrated are the FeSe stacking configurations where the Fe atoms are aligned on top of the TiO$_x$ site with the largest charge distribution change.

Although light-induced persistent structural changes in SrTiO$_3$ are very weak, we have found previously [16] that, UV excitation can produce a volatile metal-insulator transition in bare SrTiO$_3$ crystal with a TiO$_x$-TiO$_2$ termination at the < 40 K low temperature range where the competition between antiferrodistortive and ferroelectric instabilities in SrTiO$_3$ is well-known.[29-32] Such an effect indicates a significant photo-modification of the electron density in the SrTiO3 surface layers, which can affect the structure of the FeSe monolayer through interface coupling. Density functional theory (DFT) computational results (Figure 5c and Figure S4, Supporting Information) find that such modification depends strongly on the TiO$_x$ surface structure. For the TiO$_x$ ($x$ = 1) type of surface termination, the photodoping induced change in the charge distribution is most significant near the Ti atoms at the surface and strongly protruding from of the surface. For the TiO$_x$ ($x$ = 2) type of termination, the charge distribution change is instead largest near the surface O atoms and more extended into the bulk. Such distinction may explain why the light-induced atomic displacements observed in monolayer FeSe/SrTiO3 are domain-specific. As illustrated in Figure 5c, when the stacking configuration in Domain-2 is matched with the TiO$_x$ ($x$ = 1) type of substrate termination, Fe atoms can be aligned directly on top of the surface sites where significant out-of-plane charge distribution changes take place, possibly resulting in the large out-of-plane structural corrugation observed. In contrast, Fe atoms in Domain-1 can be aligned on top of the charge modification hotspots only when matching with TiO$_x$ ($x$ = 2) type of SrTiO3 termination, leading to the in-plane distortion detected. After the photoexcitation ends, the strong interaction with the bulk layers prevents major structural change in the TiO$_x$-TiO2 double layer but forms in the more susceptible FeSe monolayer a metastable structure modification that is very difficult to synthesize directly.

3. **Discussion and Conclusion**

To summarize, owing to the CTR-COBRA technique, two low-temperature metastable structures of monolayer FeSe are discovered, produced by 3.4 eV weak continuous-wave lights with the aid of the interface interactions. These two structures have very different atomic distortions in the Fe layer and are formed from two coexisting FeSe-SrTiO3 stacking configurations. The transition of FeSe into these metastable structures likely plays a major role in the optical switching of the superconducting $T_C$ observed previously in transport measurements.[16] Firstly, the light-induced relative displacements between the Fe and Se atoms can greatly impact the electron correlation strength.[33-37] The variation of the Se-Fe-Se bonding angle closely correlates with the spectral weights of different Fe $3d$ orbitals,[33] which may contribute to the Cooper pairing differently.[38-39] Such angle-dependent electron correlation has been found to strongly increase the strength of the electron-phonon coupling in FeSe multilayers,[25] and similar enhancement could be more significant in the single-layer limit. Secondly, the altered Fe-O distance at the interface can directly affect the interfacial electron-phonon coupling as well,[40-43] especially the coupling with the surface phonons formed from the out-of-plane vibrations of the O atoms in the $TiO_x$-$TiO_2$ double layer.[27, 44] Thirdly, the metastable structures with changed structural polarizations can also modify the charge transfer across the interface.[45-49] Overall, the data reveals two structural traits, namely the out-of-plane corrugation of the Fe plane and the in-plane antiferrodistortive rotation of the Fe-trapezoid, that are not considered in the past but can strongly influence the electron pairing in FeSe. Further investigations along this line can help us better identify interface interaction pathways most relevant to the superconductivity enhancement in FeSe/SrTiO$_3$.

On the other hand, our study also shows that the epitaxial FeSe-SrTiO$_3$ interface is very complex with multiple stacking domains and different oxygen arrangements in the topmost $TiO_x$ layer. The different combinations of these two variations respond to the photoexcitation very differently and may also exhibit distinct superconducting transition temperatures. It is highly desirable to explore ways in the future to identify these configurations spatially and characterize their transport properties individually.

## 4. Experimental Section/Methods

*Sample preparation:* The insulating single crystal SrTiO$_3$ (001) substrates were prepared following the recipe in Ref [16]. Single-layer FeSe films were grown at ~ 400 °C for 10 minutes with a flux ratio of Fe to Se 1: 10 ~ 20. The as-grown films were then annealed at ~ 550 °C for 1 hour. For *ex-situ* measurements, single-layer FeSe/SrTiO$_3$ samples were protected by ~ 15 quintuple layers of Sb$_2$Te$_3$ films grown at ~ 200 °C for 45 minutes by co-evaporating Sb and Te with a flux ratio of ~ 1: 5.

*STM:* Measurements were carried out at 4.5 K using a Unisoku ultrahigh vacuum low-temperature STM. A polycrystalline PtIr tip was used and tested on Ag/Si(111) films before the STM measurements.

*ARPES:* Measurements were carried out with a Scienta DA30 analyzer and He discharge lamp (h$\nu$ = 21.218 eV). The energy resolution was set at 10 meV and the angular resolution is 0.3°. The Fermi level was determined by measuring the Ag film on Si (111) substrate.

*CTR measurement and COBRA analysis:* Synchrotron X-ray surface diffraction crystal truncation rod (CTR) measurements were performed on a Newport Kappa six-circle diffractometer using an X-ray energy of 20 KeV at sector 33-ID-D, E of the Advanced Photon Source, Argonne National Laboratory. The total X-ray flux is about the $2.0 \times 10^{12}$ photons s$^{-1}$. The X-ray beam was focused by a pair of Kirkpatrick–Baez mirrors down to a beam profile of ~80 μm (vertical) × 200 μm (horizontal). The two-dimensional diffraction images of CTRs at the out-of-plane L direction in the reciprocal space were taken by a pixel array area detector (Dectris PILATUS 100 K). Three-dimensional total electron densities for the complete atomic structures of the thin film system were reconstructed from the complete set of CTRs by using an iterative phase retrieval technique, known as coherent Bragg rods analysis (COBRA). The low temperature control used Advanced Research System cryostat compatible with Newport X-ray diffractometer to cool down the sample close to 10 K. A compact 3.4 eV UV LED module was integrated with the cryostat sample post via its electric feedthrough to external DC power supply.

*DFT+DMFT computations:* We use the embedded implementation of the all-electron DFT+dynamical mean field theory (DMFT) method, where the self-energy is approximated by a quantity local to the atom (or cluster) in the unit cell. We optimize the atom positions in DFT+eDMFT functional,[50-51] which is defined in real space and implemented in the very accurate all-electron linear augmented plane wave (LAPW) basis, as implemented in Wien2k [52] The quantum impurity method is solved by the numerically exact continuous-time quantum Monte Carlo method (CTQMC).[53] The charge density is calculated self-consistently and the DFT exchange-correlation energy is evaluated on this self-consistent density to obey the stationary of the Klein functional. All five Fe *d*-orbitals were treated as correlated, and states within ±10 eV of the Fermi level were included in hybridization. The values of Hubbard $U$ and Hund's $J$ were set to 5 and 0.7 eV, respectively, consistent with other studies of Fe-based superconductors using this code.[33, 54] This code was previously used to study a variety of strongly correlated materials with transition metal compounds.[55-56]

*DFT photoresponse computations:* We construct a SrTiO$_3$ slab containing 5 layers of SrO/TiO$_2$ with TiO$_x$ ($x$ = 1, 2) termination to model the photoresponses measured in the experiments, employing the cubic lattice structure to start. DFT calculations were performed using the VASP software package,[57-58] employing the Projector Augmented-Wave (PAW) method for accurate electron-ion interactions.[59] The exchange-correlation effects were treated within the Generalized Gradient Approximation (GGA) using the Perdew-Burke-Ernzerhof (PBE) functional.[60] A plane-wave energy cutoff of 560 eV was used to ensure convergence, and the Brillouin zone was sampled using a Monkhorst-Pack grid with a k-point mesh of 12 × 12 × 3. We fix the position of Ti atoms based on the COBRA-determined structure, and fully relax the O atoms until the forces on each atom were less than 0.02 eV/Å, and the total energy convergence criterion was set to $10^{-6}$ eV.

**Acknowledgements**

CTR measurements at ANL is supported by US Department of Energy Grant No. DE‐SC0021393 and ANL Beamtime is granted under user proposal GUP-72989. Work performed by C. C.'s group is supported by National Natural Science Foundation of China Award No. 12274450, Chinese Academy of Sciences Award No. YSBR-100, and Ministry of Science and Technology of People's Republic of China Award 2022YFA1403000. N. W. and Y. W acknowledge funding support from National Natural Science Foundation of China (No. 12474246). Work performed by L. L. and M.W.'s groups are supported by the U.S. Department of Energy, Office of Basic Energy Sciences, Award Nos. DE-SC0017632 and DE-SC0021393.

**Data Availability Statement**

The data can be provided by CC pending scientific review and a completed material transfer agreement. Requests for the data should be submitted to: cheng.cen@iphy.ac.cn.